\documentstyle[11pt]{article}
\def\lsim{\mathrel{\rlap{\raise 2.5pt \hbox{$<$}}\lower 2.5pt}}
\def\gsim{\mathrel{\rlap{\raise 2.5pt \hbox{$>$}}\lower 2.5pt}}
\begin{document}
\bibliographystyle{plain}
\thispagestyle{empty}
\begin{small}
\begin{flushright}
IISc-CTS-1/97\\
BUTP-97/4\\ 
hep-ph/9702279\\
\end{flushright}
\end{small}
\vspace{-3mm}
\begin{center}
{\Large
{\bf
Status of Supersymmetric Grand Unified Theories}}
\vskip 0.5cm
B. Ananthanarayan\footnote{Invited Speaker at the XII DAE Symposium on
High Energy Physics, Guwahati, India, Dec. 26, 1996 - Jan. 1, 1997}\\
Centre for Theoretical Studies, \\
Indian Institute of Science, \\
Bangalore 560 012, India.\\

\medskip

P. Minkowski\\

Institut f\"ur Theoretische Physik,\\
 Universit\"at Bern, 5 Sidlerstrasse,\\
CH 3012, Bern, Switzerland.\\
\vskip 1cm
To the Memory of Prof. Abdus Salam
\vskip 1cm
\end{center}
\begin{abstract}
We begin with a brief discussion of the building blocks of
supersymmetric grand unified theories. 
We recall some of the compelling theoretical reasons
for viewing supersymmetric grand unification as an attractive
avenue for physics beyond the standard model.  This is
followed by a
discussion of some of the circumstantial evidence
for these ideas.  
\end{abstract}

\newpage

\section{Introduction}
The standard model of the strong and electro-weak interactions is
based on a Lagrangian field theory of quark, leptonic, scalar Higgs
and gauge bosonic degrees of freedom\cite{gsw,cl}.  Central to the standard
model are the principles of gauge invariance and its spontaneous
symmetry breakdown via the Higgs mechanism.  The standard model
predicts the existence of a scalar Higgs particle which is the
remnant of the Higgs mechanism by which the gauge bosons of
the broken generators of $SU(2)_L\times U(1)_Y$ become massive
when the gauge symmetry is broken down to the residual $U(1)_{em}$.
The mass of the yet to be discovered Higgs boson is
not fixed therein, but is bounded from below from present day experiments
and from above by requirements of vacuum stability.  

Indeed, if the
standard model were to be vindicated by the discovery of the Higgs
grand unification appears to be a path to go beyond the energies
where the standard model is the correct theory, while continuing
to be based on these principles.  There would then be
a unification scale $M_G\sim 10^{16}
{\rm GeV}$ suggested by gauge coupling unification,
above which physics would be described
by a grand unified theory\cite{ggr}
 based on a gauge group $G$.   Such a theory
would then make a whole host of predictions and simplifications
of our understanding of fundamental phenomena.  
A compelling goal of theoretical physics
is to replace what are the
engineering aspects of the standard model by a fundamental theory;
for example
arbitrary parameters of the standard model, hitherto fixed
by experiment,  would then
be explained as consequences of a unified and symmetric
structure.  Furthermore, within grand unified theories, one uncovers
highly desirable properties such as anomaly free representations of
certain grand unified groups. 
One expects the unification of hitherto unrelated quantum
numbers such as baryon and lepton numbers.   These in turn imply
concrete low-energy predictions which can be confronted by experimental
and/or observational information.  

The presence
of disparate scales in the theory, $M_G$ and the weak scale $M_W
\sim 174\ {\rm GeV}$,
expected to be separated by more than ten orders of magnitude,
would render the mass of the Higgs scalar of the electro-weak model
$\sim M_W$, unnatural-natural.  Should the Higgs scalar be elementary, then
one manner in which it would remain naturally at the weak scale is due
to cancellation of divergences as in 
supersymmetric unified models\cite{mj,hpn}.
Supersymmetry\cite{ms}
is the only symmetry that has non-trivial commutation
relations with the generators of the Lorentz group and is a fermionic
object that interchanges bosonic and fermionic degrees of freedom.
Although supersymmetry does not appear to be manifest, it could be
broken softly while preserving 
all the desirable properties of supersymmetric
theories.  Models with softly broken susy are popular and 
significant experimental effort will be made 
to test the predictions of these models.  

The final frontier that still remains to be explored
is a framework within which a consistent incorporation 
of the gravitational
interactions is successful.  Whereas it has not been
possible to replace the Einstein theory 
by a quantum version due to bad ultra-violet
behaviour, supergravity possesses improved ultra-violet 
properties\cite{mj}.
String theories\cite{gsw2}
often contain supergravity in their low energy spectrum
and as a result supersymmetric unification is a favored candidate for
these reasons as well.

Other significant avenues exist for the exploration of these
theories.  Note, for instance, non-perturbative aspects of the
theory such as the possibility of finding topological defects 
at the time of spontaneous symmetry breakdown when combined with
standard big bang cosmology imply specific constraints on grand
unified models. 
Examples of such defects are monopoles, cosmic strings
and domain walls.  

The task of this talk is to briefly summarize the building blocks
of supersymmetric grand unification and recall 
the main circumstantial evidence for the program.
The most significant advance from the experimental 
direction has come with the precision measurements of
the gauge coupling constants at the $Z^0$ factory
LEP\cite{somganguly}
 (and SLC) and the discovery of the top-quark at the
Tevatron\cite{meenakshi} by the CDF collaboration and confirmed by the
D0 collaboration.  These advances place significant
constraints on scenarios of unification for a start.
More spectacular is the fact that certain scenarios of
unification predicted that the top-quark mass would
have to be sufficiently large and roughly in the range
where it has been found.
Note that combinations of theoretical tools such
as the requirement of infra-red fixed point structure
of Yukawa couplings as well as finiteness also accommodate
top-quark masses in this range. 
Challenges today lie in 
spotting the first traces of the supersymmetric partners
of the known particles, e.g.
figuring out search strategies for these
for future collider experiments as well as at non-accelerator
experiments.

\section{Spontaneously broken gauge theories}
Whereas the gauge invariance
of the standard model rests of the gauge group $SU(3)_C\times
SU(2)_L\times U(1)_Y$, with the quark, lepton [matter] fields
and Higgs fields transforming in a specific manner under the
gauge group, at low energies, the $SU(2)_L\times U(1)_Y$ is
spontaneously broken to the $U(1)_{em}$ subgroup at the weak scale
via the Higgs mechanism.  The result is that three
of the gauge bosons, $W^{\pm}$ and $Z^0$ pick up masses at the
weak scale as does the neutral Higgs scalar.  The fermions become
massive through the Yukawa couplings to the scalars since the
vacuum expectation value $<\phi>\neq 0$.

It is possible to envisage a scenario
wherein this is embedded in a larger group $G$, which would be
the basis of the gauge invariance of a theory manifest above
a unification scale $M_G$, below which it would be spontaneously
broken via the Higgs and possibility
some other mechanism to a sub-group large enough to
contain the standard model (in a multi-step scenario), which would
then be further broken down to the standard model gauge group
at various stages.  

Circumstantial evidence for this, 
is found from the renormalization
group evolution of the gauge coupling constants of the standard
model gauge group, which appears to bring them all together at
a large scale $M_G\sim 10^{16}\ {\rm GeV}$ when {\it the normalization
on the hyper-charge coupling constant as required by grand unification
is imposed.}

Indeed, the arrival at the structure
of fundamental interactions from renormalization group flow
has a predecessor in the example of asymptotic freedom in
deep inelastic scattering experiments and thus gauge
coupling unification is an extremely encouraging sign that
grand unified theories are the right step for a theory
of fundamental interactions.    
Earliest examples of grand unification were provided by
those based on the groups $SU(4)\times SU(2)\times SU(2)$,
$SU(5)$ and $SO(10)$.  

Grand unification, of course, implies more than just the
coming together of the gauge coupling constants.  One would
be gratified if it were possible to unify the particle
content of the theory as well.  Indeed,
simplifying features of grand unification include
embedding several of the matter fields of the standard model into
irreducible representations of the underlying gauge group.
That such an embedding should at all be possible is an
astonishing property of grand unified theory:  furthermore,
it has the capacity to explain the charge ratios for the
elementary fermions in terms of simple group theory.
It turns out that $SO(10)$\cite{fm}, for
instance, still remains one of the most elegant unification
groups, with an entire standard model family and a right handed
neutrino accommodated in a single 16 dimensional representation.

Whereas grand unified theories are based on local Lagrangian
field theories possessing symmetries, it is then important to
address the question of anomalies in such theories.  Compelling
theoretical reasons for viewing grand unification as a consistent
road to physics beyond the standard model include the fact that
several grand unified groups ensure the vanishing of anomalies of
gauge currents from the very nature of their representations; e.g.,
for any irreducible representation
of $SO(10)$, $Tr (Y)$ and $Tr (Y^3)$ vanish automatically\cite{ggr},
where $Y$ is the hypercharge generator.
Thus anomaly cancellation which may appear somewhat
mysterious in the standard model is natural in grand
unification;  it may be worth noting that while the structure
of the strong interaction was arrived at through the analysis
of the anomaly in $\pi^0 \rightarrow 2 \gamma$\cite{cl}, the structure
of theories beyond the standard model may also be uncovered
from such considerations of anomalies. 
In addition, global anomalies are related to the centre of the gauge
group:  $Z_2$ in the case of $SU(2)$, $SO(10)$ and $Z_3$ in
$E_6$.

In turn, processes involving transitions from one set of matter
fields to another predicting, say the decay of the proton at measurable
rates are intrinsic features of unification.  
The continued failure of the proton to decay
within present day experiments in turn implies
constraints on scenarios of grand unification\cite{hm}.

Note that whereas in the standard model, the field content
forbids a Dirac mass for the neutrinos since the right
handed neutrino is absent and Majorana mass is forbidden
by the conservation of lepton number.  In grand unified models,
neither of these principles is respected and a wide variety
of possibilities exists for the generation of neutrino masses.
However, far from being arbitrary, it should be possible to
uncover information regarding the structure of unified theories
from accurate determination of
small and eventually 
large neutrino masses and mixing angles,
{\it viz.}, neutrino masses may be viewed as bearing an imprint
on the structure of grand unification and the nature of the
breakdown of unification\cite{pm}.

\section{Supersymmetric unification}
Supersymmetry is the unique symmetry that has non-trivial
commutation relations with the generators of the Lorentz group.
Supersymmetries enjoy non-trivial anti-commutation relations
amongst each other.  Their action on representations of the
supersymmetry algebra interchange the statistics between the
members.  Linear representations of the supersymmetry
algebra in relativistic field theory are realized in the Wess-Zumino
model\cite{ms}.   Important representations include chiral multiplets and
vector multiplets, which form the basis of the extension of the
standard model to various supersymmetric versions of the standard
model.  Since supersymmetry is not manifest in nature,
it must be broken, either spontaneously or explicitly.  It appears
that the second option is more favored, certainly more popular,
wherein supersymmetry is broken explicitly but softly.  The
requirement of soft supersymmetry breaking is in accordance with
the requirement of the well-known properties of supersymmetric
models including the cancellation of quadratic mass divergences for
scalars.  

In the context of grand unified model building, the existence of
scales $M_W$ and $M_G$ separated by several orders of magnitude
renders the mass of the elementary Higgs of the standard model
unstable and would drive it to the unification scale, without an
un-natural fine tuning of parameters of the Lagrangian.  The
cancellation of quadratic divergences in manifestly and softly-broken
supersymmetric theories renders supersymmetric versions of grand
unified models attractive candidates for unification.
The program of writing down a supersymmetric version of the
standard model, which is then embedded in a grand unified scheme,
[alternatively a supersymmetric version of a grand unified scheme]
may be realized by replacing every matter and Higgs field, by
a chiral superfield whose members carry the same gauge quantum numbers,
and by replacing every gauge field, by a vector super-multiplet.
Supersymmetry also requires that the standard model Higgs
doublet is replaced by two Higgs multiplets.  
This in turn leads to the introduction of another parameter
$\tan\beta$ which is defined as the ratio of the vacuum expectation
values of these two Higgs fields, $v_2/v_1$ where $v_2$ and $v_1$ 
are the vacuum expectation value of the Higgs fields that provide
the mass for the up-type quark and the down-type and charged leptons
respectively.
All the interactions
of the resulting model may then be written down once the superpotential
is specified.  Note that gauge invariance and supersymmetry allow
the existence of a large number of couplings in the effective theory
that would lead to proton decay at unacceptably large rates.
An ad hoc symmetry called R-parity is imposed on the resulting model
which eliminates these undesirable couplings and such a version has
received the greatest attention for supersymmetry search.
More recently models have been and are being considered where R-parity
is partially broken in order to study the implications to collider
searches.  
However such models are constrained by bounds on flavor
changing neutral currents as well as by the standard CKM picture,
also as it applies to CP violating phases.

In what follows we recall some of the 
essential successes of the recent investigations\cite{review} in
the theory of supersymmetric unification.
This was spurred by the confrontation of the ideas
of unification by the precision measurements of
the gauge couplings of the standard model at the
LEP\cite{adf}.  A highly simplified understanding of this
feature may be obtained from a glance at the one-loop
evolution equation for the standard model gauge couplings,
more correctly the gauge couplings of the minimal
supersymmetric standard model assuming that the effective
supersymmetry scale is that of the weak scale, with
$t=\ln \mu$:
$
\frac{d\alpha_i}{dt}=\frac{\alpha_i^2}{2 \pi} b_i, 
\ b_1=33/5,\ b_2=1,\ b_3=-3,
$
where we have assumed three generations.  One may then integrate
these equations to obtain:
$
\frac{1}{\alpha_i(M_Z)}=\frac{1}{\alpha_i(M_G)}+\frac{b_i}{2\pi}\ln
\frac{M_G}{M_Z}.
$
One may then use the accurately known value of $\alpha_{em}
(M_Z)=1/128$, with the identity $1/\alpha_{em}=5/3\alpha_1+1/\alpha_2$
which accounts for the normalization imposed by unification,
and the values of $\alpha_3(M_Z)\approx 0.12$ to solve for the
unification scale $M_G$ and the unified coupling constant $\alpha_G
\equiv 
\alpha_{1,2,3}(M_G)$.  One then has a prediction for $\sin^2\theta_w$
at the weak scale which comes out in the experimentally measured range.
Sophisticated analysis around this highly simplified picture up to
two and even three loops taking into account the Yukawa couplings
of the heaviest generation which contribute non-trivially at the
higher orders, threshold effects, etc., vindicate this picture of
gauge coupling unification which today provides one of the strongest
pieces of circumstantial evidence for grand unification\cite{deBoer}.

Predictions arising from (supersymmetric) unification
such as for the mass of the top-quark 
have been vindicated experimentally.
It turns out that unification based on $SO(10)$ 
is a scheme with great predictive power not merely in the
context of top-quark mass but also with implications
for the rest of the superparticle spectrum.  The primary requirement
that is imposed is that the heaviest generation receives its mass
from a unique coupling in the superpotential
$h{\bf 16.16.10}$
where the {\bf 16} contains a complete generation and
the complex {\bf 10} the two electroweak doublets\cite{als}.  When
the Yukawa couplings of the top and b-quarks and the $\tau$-lepton
are evolved down to the low energy and $\tan\beta$ pinned
down from the accurately known $\tau$-mass, one has a unique
prediction for the b and top-quark masses for a given value
of $h$.  If $h$ is chosen so as to yield $m_b(m_b)$ in its
experimental range, the top-quark mass is uniquely determined
up to these uncertainties.  
Now $\tan\beta\simeq m_t/m_b$ and the top-b hierarchy
is elegantly explained in terms of this ratio coming out
large naturally.
 
It is truly intriguing that
this picture yields a top-quark mass 
in its experimental range,
with $\alpha_S$ in
the range of the LEP measurements 
despite the complex interplay between the evolution equations
involved, the determination of the unification scale, running
of QCD couplings below the weak scale.  Note that this
requires that the top-Yukawa coupling must also come out
of order unity at $M_Z$.  It is also worth noting that
due to the nature of the evolution equations and competition
between the contributions to these from the gauge and Yukawa
couplings, this number $m_t(m_t)$ lies near a quasi-fixed point of its
evolution, {\it viz}, there is some insensitivity to the
initial choice of $h$\cite{sw}.  Moreover, if the $SO(10)$ unification
condition is relaxed to an $SU(5)$ one where only the b-quark
and $\tau$-lepton Yukawa couplings are required to unify at
$M_G$, $m_t(m_t)$ comes out in the experimental range
while preserving $m_b(m_b)$ in its experimental range for
$\tan\beta$ near unity.  In this event also the top-quark
Yukawa coupling lies near a quasi-fixed point which is numerically
larger compensating for the smaller value of $\sin\beta$
that enters the expression for its mass: $m_t=h_t \sin\beta 174
\ {\rm GeV}$.  Another interesting connection arises in
this context between the values of the Yukawa couplings at
unification and that of the gauge coupling when one-loop
finiteness and reduction of couplings is required:  such
a program also yields top-quark masses in the experimental
range\cite{kubo}.  

Besides the vindication of top-quark discovery predicted
by susy guts, another strong test takes shape in the form
of its prediction for the scalar spectrum.  In the MSSM 
the mass of the
lightest scalar is bounded at tree level by $M_Z$ since
all quartic couplings arise from the D-term in the scalar
potential.  The presence of the heavy top-quark enhances
the tree-level mass, but the upper bound in these models
is no larger than $140\ {\rm GeV}$.

Other predictions for softly-broken susy models arise
when a detailed analysis of the evolution equations
of all the parameters of the model are performed and
the ground state carefully analyzed.  In the predictive
scheme with $SO(10)$ unification, the model is further
specified by $M_{1/2},\ m_0$ and $A$, the common
gaugino, scalar and tri-linear soft parameters\cite{hpn}. 
It turns out that in this scheme $M_{1/2}$ is required
to come out to be fairly large, at least $\sim 500\ {\rm GeV}$
implying a lower bound on the gluino mass of a little
more than a TeV
and providing a natural explanation for the continuing absence of
observation of susy particles from scenarios based on radiative
electro-weak symmetry breaking\cite{als2}.
[An extensive study of the NMSSM with $SO(10)$ conditions
has also been performed\cite{ap}.]
Considerably greater freedom exists when the $SO(10)$
boundary condition is relaxed\cite{abs}.
In summary many predictions and consistency of the
MSSM and its embedding in a unified framework have been
vindicated;  however it is important to continue theoretical
investigations and checks to the consistency of these 
approaches and extensions
to include the lighter generations\cite{morereviews}.

\section{Monopoles}

This discussion is somewhat off the main stream
of the discussion above.  
Furthermore, if one were to discuss 
spontaneously broken gauge 
field theory at finite
temperatures, when temperatures reach the scale
of symmetry breaking, then phase transitions are
expected to occur which restore broken symmetries.
Indeed, at such phase transitions, one expects the
formation of topological defects which may be 
characterized by certain topological properties 
known as homotopy groups of the coset space: $G/H$,
where $G$ is the gauge group that is broken to the
subgroup $H$.    Examples of topological defects
are domain walls, strings and monopoles, which may
have been produced in the early universe as the
universe cooled to present temperatures.  
This is an example of an aspect of gauge field theory
that is outside the realm of perturbation theory.
However, certain interesting preliminary
investigations indicate that
monopoles are inconsistent as asymptotic states; they
are confined even if in the topologically parallelizable
sector the gauge theory serving as non-abelian basis
to the classically acceptable monopole solutions 
is broken\cite{striebel}.
[Other examples of standard model physics that lie
outside this realm is that of the formation of
fermion condensates that are required to spontaneously
break chiral symmetry that lead to the generation of
massless pions when the quark masses are set to zero.]

\medskip

\noindent{\bf Acknowledgments:}  BA thanks G. Zoupanos for
discussions.


\begin{thebibliography}{abcdefg}

\bibitem{gsw} 
S. Glashow,  Nucl. Phys. 22 (1961) 579; S. Weinberg, Phys. Rev. Lett.
19 (1967) 1264; A. Salam in {\it Elementary Particle Theory},
ed. N. Svartholm, Almqvist and Wilsell, Stockholm, 1969, p. 367.

\bibitem{cl}
For a comprehensive
discussion see, e.g., T-P. Cheng and L-F. Li,
{\it Gauge theory of elementary particle physics},
Clarendon Press, Oxford, 1984.

\bibitem{ggr} See, e.g., G. G. Ross, {\it Grand Unified 
Theories}, The Benjamin/Cummings Publishing Company, Inc.,
Menlo Park, California, 1985.

\bibitem{mj} For a collection of reports, see M. Jacob, ed.,
{\it Supersymmetry and Supergravity, A Reprint Volume of
Physics Reports}, North-Holland/World Scientific, Amsterdam/Singapore,
1986.

\bibitem{hpn} H-P. Nilles, Phys. Rep. 110 (1984) 1, reprinted
in Ref.\cite{mj}.

\bibitem{ms} See, e.g., M. Sohnius, Phys. Rep. 128 (1985) 39,
reprinted in Ref.\cite{mj}. 

\bibitem{gsw2} See, e.g., M. Green, J. Schwarz and E. Witten,
{\it Superstring Theory:1, 2},
Cambridge University Press, Cambridge, 1987.


\bibitem{somganguly} S. Ganguly, Invited talk at this conference.

\bibitem{meenakshi} M. Narayan, Invited talk at this conference.

\bibitem{fm} H. Fritzsch and P. Minkowski, Ann. Phys. 93 
(1975) 193.

\bibitem{hm} For a recent review, see e.g., H. Murayama, 
{\it Nucleon decay in GUT and nonGUT SUSY models},
hep-ph/9610419.

\bibitem{pm} For a recent review, see e.g, P. Minkowski,
{\it  Neutrino mass and mixing}, Bern University preprint, BUTP-95/22.

\bibitem{review} For some recent reviews, see, e.g.,
L. J. Hall, {\it The heavy top-quark and supersymmetry}, hep-ph/9605258;
F. Zwirner, {\it Extensions of the standard model}, hep-ph/9601300;
S. Pokorski,{\it Status of the minimal supersymmetric standard
model}, hep-ph/9510224; S. Dimopoulos, {\it Beyond the standard
model}, ICHEP 1994: 93-106 (QCD161: H51: 1994).

\bibitem{adf} U. Amaldi, W. de Boer and H. F\"urstenau,
Phys. Lett. B 260 (1991) 447; P. Langacker and M-X. Luo,
Phys. Rev. D 44 (1991) 817; 
C. Giunti, C. W. Kim and U. W. Lee, Mod. Phys. Lett. A6
(1991) 1745.

\bibitem{deBoer} For updates, see e.g., W. de Boer, 
{\it The constrained MSSM revisited}, hep-ph/9611394;
{\it Global fits to the MSSM and SM to electroweak
precision data},
hep-ph/9611395.

\bibitem{als} B. Ananthanarayan, G. Lazarides and
Q. Shafi, Phys. Rev. D 44 (1991) 1613; For a recent
update see, U. Sarid, {\it Precision top mass 
measurements vs. Yukawa unification predictions}, hep-ph/9601300.

\bibitem{sw} For a review, see, e.g., B. Schrempp and
M. Wimmer, {\it Top quark and Higgs boson masses:
interplay between infrared and ultraviolet physics}, hep-ph/9606386.

\bibitem{kubo} J. Kubo, M. Mondrag\'on and G. Zoupanos,
{\it Top quark mass predictions from gauge-Yukawa unification},
hep-ph/9512400.

\bibitem{als2} B. Ananthanarayan, G. Lazarides and Q. Shafi,
Phys. Lett. B300 (1993) 245; B. Ananthanarayan, Q. Shafi
and X-M. Wang, Phys. Rev. D50 (1994) 5980 and references therein.

\bibitem{ap} For a comprehensive
analysis of the NMSSM with large
$\tan\beta$, see
B. Ananthanarayan and P. N. Pandita, Phys. Lett.
B 353 (1995) 70; Phys. Lett. B 371 (1996) 245.

\bibitem{abs} B. Ananthanarayan, K. S. Babu and Q. Shafi,
Nucl. Phys. B 428 (1994) 19 and references therein.

\bibitem{morereviews} For other recent directions see, e.g.,
T. Bla$\check{z}$ek et al., {\it A global $\chi^2$ analysis 
of electroweak data in SO(10) SUSY GUTs},
hep-ph/9611217; M. Carena et al., {\it Bottom-up approach and 
SUSY breaking}, hep-ph/9610341.

\bibitem{striebel} 
M. Striebel, 
{\it Magnetic Monopoles in a Constant Background Gauge Field,}
             University of Bern thesis,  1987 (unpublished)

\end{thebibliography}
\end{document}